\input pipi.sty
\input epsf.sty
\input psfig.sty
\magnification1054


\nopagenumbers
\rightline{FTUAM 05-1}
\rightline{January, 13,  2005}
\bigskip
\hrule height .3mm
\vskip.6cm
\centerline{{\bigfib The  
 scalar radius of the pion}}
\medskip
\centerrule{.7cm}
\vskip1cm

\setbox9=\vbox{\hsize65mm {\noindent\fib F. J. 
Yndur\'ain} 
\vskip .1cm
\noindent{\addressfont Departamento de F\'{\i}sica Te\'orica, C-XI\hb
 Universidad Aut\'onoma de Madrid,\hb
 Canto Blanco,\hb
E-28049, Madrid, Spain.}\hb}
\smallskip
\centerline{\box9}
\bigskip
\setbox0=\vbox{\abstracttype{Abstract} The pion  scalar radius  
 is given by $\langle r^2_S\rangle=(6/\pi)\int_{4M^2_\pi}^\infty{\rm d}s\,\delta_S(s)/s^2$, 
with $\delta_S$  the phase of the scalar form factor. 
Below $\bar{K}K$ threshold, $\delta_S=\delta_\pi$,  $\delta_\pi$ being the isoscalar, 
S-wave $\pi\pi$ phase shift. At high energy, $s>2\,{\rm GeV}^2$, $\delta_S$ is given by 
perturbative QCD. In between I  argued, in a previous letter, 
that one can interpolate   $\delta_S\sim\delta_\pi$, because 
inelasticity is small, compared with the errors. 
This gives $\langle r^2_S\rangle=0.75\pm0.07\,{\rm fm}^2$. 
Recently, Ananthanarayan, Caprini, Colangelo, Gasser and Leutwyler (ACCGL) have claimed that this
 is  incorrect and one should have instead $\delta_S\simeq\delta_\pi-\pi$; 
then $\langle r^2_S\rangle=0.61\pm0.04\,{\rm fm}^2$.
Here I show that the ACCGL phase $\delta_S$ is pathological in that it is discontinuous for small 
inelasticity, does not coincide with what perturbative QCD suggests at high energy, and only occurs
 because  these authors take a value for $\delta_\pi(4m^2_K)$ different from  what
 experiment indicates. If one uses the  value for $\delta_\pi(4m^2_K)$ favoured by experiment,
 the ensuing phase $\delta_S$  is continuous,  agrees with perturbative QCD expectations,
 and satisfies  $\delta_S\simeq\delta_\pi$, thus confirming the correctness of my previous estimate,
 $\langle r^2_S\rangle=0.75\pm0.07\,{\rm fm}^2$. 
}
\centerline{\box0}
\brochureendcover{Typeset with \physmatex}
\brochureb{\smallsc f. j.  yndur\'ain}{\smallsc the  
 scalar radius of the pion}{1}

\booksection{1. Introduction}

\noindent
The quadratic scalar radius of the pion, $\langle r^2_{{\rm S}}\rangle$, is defined via the 
scalar form factor, $F_{S,\pi}$:
$$
F_{S,\pi}(t)\simeqsub_{t\to0}F_{S,\pi}(0)\Big\{1+\tfrac{1}{6}\langle r^2_{{\rm S}}\rangle\,t\Big\}, 
\equn{(1.1)}$$
where
 $$\langle\pi(p)|\big[m_u\bar{u}u(0)+m_d\bar{d}d(0)\big]|\pi(p')\rangle=(2\pi)^{-3}F_{S,\pi}(t);
\equn{(1.2)}$$
 the (charged) pion states are normalized to $\langle \pi(p)|\pi(p')\rangle=
2p_0\delta({\bf p}-{\bf p}')$, and $t=(p-p')^2$.
To one loop in chiral perturbation theory (ch.p.t.), 
$\langle r^2_{{\rm S}}\rangle$ is related to the important coupling constant $\bar{l}_4$ by
$$\langle r^2_{{\rm S}}\rangle=\dfrac{3}{8\pi^2f^2_\pi}\Big\{\bar{l}_4-\tfrac{13}{12}\Big\}.
\equn{(1.3)}$$
$f_\pi\simeq93\,\mev$ is the decay constant of the pion. 

An evaluation of $\langle r^2_{{\rm S}}\rangle $ was given some time ago by 
Donoghue, Gasser and Leutwyler;\ref{1} we will refer to this paper as DGL. 
These authors found (we quote the improved result from the second paper in ref.~1)
$$\langle r^2_{{\rm S}}\rangle_{\rm DGL}=0.61\pm0.04\;{\rm fm}^2,\quad
\bar{l}_4=4.4\pm0.2.
\equn{(1.4)}$$
The error comes from experimental errors and the estimated higher order corrections.

As noted in ref.~2,  one can obtain 
the scalar radius from the sum rule
$$\langle r^2_S\rangle=\dfrac{6}{\pi}\int_{4M^2_\pi}^\infty\dd s\,\dfrac{\delta_S(s)}{s^2},
\equn{(1.5)}$$
where $\delta_S(s) $ is the phase of $F_{S,\pi}(s)$, and $M_\pi$ is the charged 
pion mass. At low energy, $\delta_S(s)=\delta_\pi(s)$, where 
$\delta_\pi(s)$ is the phase shift for $\pi\pi$ scattering  with isospin zero in the S wave.
This equality holds with good accuracy up to the opening of the $\bar{K}K$ threshold, 
at $s=4m^2_K$; for $m_K$ we take the average kaon mass, $m_K=496\,\mev$. 
At high energy, $s>2\,\gev^2$, one can use the asymptotic estimate that 
perturbative QCD indicates for $\delta_S(s)$ (see below) and, between these two regions, 
what was considered in ref.~2
a reasonable interpolation, viz., $\delta_S(s)\sim\delta_\pi(s)$. 
One then finds,
$$\langle r^2_{{\rm S}}\rangle=0.75\pm0.07\;{\rm fm}^2,\quad
\bar{l}_4=5.4\pm0.5.
\equn{(1.6)}$$
This is  about $2\,\sigma$ above the DGL value, Eq.~(1.4).

The integral in (1.5) up to $s=4m^2_K$ can be evaluated in a fairly unambiguous manner, 
and  the contribution of the high energy region, $s>2\,\gev^2$, although evaluated with 
different methods, is found similar in refs.~1,~2,~3. 
The conflictive contribution is that of the intermediate region,
$$\int_{4m^2_K}^{2\,{\gev}^2}\dd s\,\dfrac{\delta_S(s)}{s^2}.
\equn{(1.7)}$$
In fact, very recently Ananthanarayan, Caprini, Colangelo, Gasser and Leutwyler,\ref{3}
that we will denote by ACCGL, have challenged the result of ref.~2. 
Their main objection is that the Fermi--Watson 
final state interaction theorem does {\sl not} guarantee 
that $\delta_\pi(s)$ and $\delta_S(s)$ are equal, even if inelasticity is negligible; 
it only requires that they differ in an integral multiple of $\pi$:
$$\delta_S(s)\simeq\delta_\pi(s)+N\pi.
\equn{(1.8)}$$
At $\pi\pi$ threshold, both $\delta_s$ and $\delta_\pi$ vanish, hence $N=0$ here.
Below  $s=4m^2_K$, continuity guarantees that the $N$ in (1.8) still vanishes, 
as assumed in ref.~2. For
 $1.7\,\gev^2\lsim s\lsim\;2\,\gev^2$ inelasticity is also 
compatible with zero. However, since this is 
separated from the low energy region by the region $2m_K< s^{1/2}\lsim1.2\,\gev$, 
where inelasticity
 is {\sl not} negligible, one can have $N\neq0$. 
Actually, ACCGL conclude that
$$\delta_S(s)\simeq\delta_\pi(s)-\pi,\quad
1.1\;\gev\simeq s^{1/2}\simeq 1.42\;\gev.
\equn{(1.9)}$$
According to ACCGL, this brings the value of $\langle r^2_{{\rm S}}\rangle $ 
back to the DGL number in (1.4).

The remark of ACCGL leading to (1.8) is correct. Nevertheless, we will 
here show that their conclusion (1.9) is wrong. 
In fact, arguments of (A) Continuity of $\delta_S(s)$ when the inelasticity goes to zero; 
(B) The value experiment indicates for the quantity  $\delta_\pi(4m^2_K)$; 
(C) The value SU(3) ch.p.t. implies for the real part of the $\bar{K}K$ scattering length 
and  $\delta_\pi(4m^2_K)$; and, 
finally, (D) The matching with the phase expected from perturbative QCD at high virtuality, 
all imply that the number $N$ in (1.8) 
vanishes, therefore substantiating  the claims of ref.~2.

It should also be noted  that the error analysis of DGL and ACCGL must be incomplete. 
With a correct error analysis, and even starting from their assumptions, DGL and ACCGL should 
have obtained a value for $\langle r^2_{{\rm S}}\rangle$ compatible 
with that in ref.~2, within errors. 
This is also discussed below.

\booksection{2. Some definitions}

\noindent
Since we will only consider the S wave for isospin zero, we will omit 
isospin and angular momentum indices. We define a matrix 
for the partial wave amplitudes for the processes
$\pi\pi\to\pi\pi,\;\pi\pi\to\bar{K}K(=\bar{K}K\to\pi\pi)$, and $\bar{K}K\to\bar{K}K$:
$${\bf f}=\pmatrix{f_{\pi\pi\to\pi\pi}&f_{\pi\pi\to\bar{K} K}\cr
 f_{\pi\pi\to\bar{K} K}&f_{\bar{K} K\to\bar{K} K}}=
\pmatrix{\dfrac{\eta\,\ee^{2\ii\delta_\pi}-1}{2\ii}& 
\tfrac{1}{2}\sqrt{1-\eta^2}\;\ee^{\ii(\delta_\pi+\delta_K)}\cr
\tfrac{1}{2}\sqrt{1-\eta^2}\;\ee^{\ii(\delta_\pi+\delta_K)}&
\dfrac{\eta\,\ee^{2\ii\delta_K}-1}{2\ii}}.
\equn{(2.1)}$$ 
Below $\bar{K}K$ threshold, the elasticity parameter is $\eta(s)=1$;
 above  $\bar{K}K$ threshold\fnote{In the present paper 
we will neglect coupling of $\pi\pi$ to 
states other than  $\bar{K}K$, for energies below $1.42\,\gev$.}
 one has the bounds $0\leq\eta\leq1$.
We will also use a K-matrix representation of $\bf f$:
$${\bf f}=\left\{{\bf Q}^{-1/2}{\bf K}^{-1}{\bf Q}^{-1/2}-\ii\right\}^{-1},
\quad {\bf Q}=\pmatrix{q_1&0\cr0&q_2}.
\equn{(2.2)}$$
$q_a$ are the momenta,
$q_1=\sqrt{s/4-M^2_\pi},\quad q_2=\sqrt{s/4-m^2_K}$.
 
We may diagonalize $\bf f$ and find the 
{\sl eigenphases}, $\delta^{(\pm)}$,
$${\bf f}={\bf C}\{{\bf g}_D-\ii\}^{-1}{\bf C}^{\rm T},
$$
$${\bf g}_D=\pmatrix{\cot \delta^{(+)}&0\cr0&\cot \delta^{(-)}},\quad
{\bf C}=\pmatrix{\cos\theta&\sin\theta\cr-\sin\theta&\cos\theta}.
\equn{(2.3a)}$$
We will define $\delta^{(+)}$ to be the eigenphase that matches $\delta_\pi$:
$\delta^{(+)}(4m^2_K)=\delta_\pi(4m^2_K)$. Then,
$$\eqalign{
\tan \delta^{(\pm)}=&\,\dfrac{T\pm\sqrt{T^2-4\deltav}}{2},\cr
\sin\theta=&\,\left\{\tfrac{1}{2}\,\dfrac{T+\sqrt{T^2-4\deltav}-
2q_1K_{11}}{+\sqrt{T^2-4\deltav}}\right\}^{1/2};\cr
T=&\,q_1K_{11}+q_2K_{22},\quad
\deltav=q_1q_2\det{\bf K}.
}
\equn{(2.3b)}$$
This holds (near $\bar{K}K$ threshold) when $K_{11}>0$. For $K_{11}<0$, the $(\pm)$ signs  
 should be exchanged in the right hand side of the expression for $\tan \delta^{(\pm)}$, 
and the square roots in the expression for $\sin\theta$ get a minus sign. 
 Near $\bar{K}K$ threshold, 
$\sin\theta$ and $\delta^{(-)}(s)$ vanish with $q_2$. 
If inelasticity were zero ($\eta=1$) the channels would 
decouple and one 
would have ${\bf C}=1$ and $\delta^{(+)}=\delta_\pi$, $\delta^{(-)}=\delta_K$.

The {\sl phase} of the $\pi\pi\to\pi\pi$ amplitude will play an important role 
in the subsequent discussion. We will actually use the phase $\phi_\pi$ 
defined by
$$f_{\pi\pi\to\pi\pi}=\cases{
+|f_{\pi\pi\to\pi\pi}|\,\ee^{\ii\phi_\pi},\quad 0\leq\phi_\pi\leq\pi;\cr
-|f_{\pi\pi\to\pi\pi}|\,\ee^{\ii\phi_\pi},\quad \pi\leq\phi_\pi\leq2\pi.
}
$$
This definition has to be adopted to agree with the standard definition of 
the phase (shift) $\delta$ for a purely  
elastic amplitude, given by $f=\sin\delta\,\ee^{\ii\delta}$, so that 
$f=\pm|f|\ee^{\ii\delta}$ with the $(\pm)$ signs as for $\phi_\pi$ above.

 Using (2.1) one gets a simple expression for the tangent of  $\phi_\pi$:
$$\tan\phi_\pi=\left\{1+\dfrac{1-\eta}{2\eta}\big(1+\cot^2\delta_\pi\big)\right\}
\tan\delta_\pi.
\equn{(2.4)}$$

For ease of reference, we also give here 
the expressions of phase shift and inelasticity in terms of the K-matrix:
$$
\tan\delta_\pi=\cases{\dfrac{q_1|q_2|\det {\bf K}+q_1K_{11}}{1+|q_2|K_{22}},\quad s\leq 4m^2_K,\cr
\eqalign{&\,\dfrac{1}{2q_1[K_{11}+q_2^2K_{22}\det {\bf K}]}\Big\{
q^2_1K^2_{11}-q^2_2K^2_{22}+q_1^2q_2^2(\det{\bf K})^2-1\cr
+&\,
\sqrt{(q^2_1K^2_{11}+q^2_2K^2_{22}+q_1^2q_2^2(\det{\bf K})^2+1)^2-
4q_1^2q_2^2K^4_{12}}\;\Big\},\quad s\geq 4m^2_K;} 
\cr}
\equn{(2.5a)}$$
$$\eta=\sqrt{\dfrac{(1+q_1q_2\det{\bf K})^2+(q_1K_{11}-q_2K_{22})^2}
{(1-q_1q_2\det{\bf K})^2+(q_1K_{11}+q_2K_{22})^2}},\quad s\geq 4m^2_K.
\equn{(2.5b)}$$

The connection with the 
scalar form factor of the pion comes about as follows. 
We form a vector $\bf F$ with $F_{S,\pi}$ and the form factor of the kaon, $F_{S,K}$, 
and define the vector ${\bf F}'$  by
$$
{\bf F}'={\bf C}^{\rm T}{\bf Q}^{1/2}{\bf F},\quad 
{\bf F}=\pmatrix{F_{S,\pi}\cr F_{S,K}\cr}.
\equn{(2.6a)}$$
Then two-channel unitarity implies that
$$\eqalign{
F_{S,\pi}=&\,q_1^{-1/2}\left\{(\cos\theta)|F'_1|\,\ee^{\ii\delta^{(+)}}+ 
(\sin\theta)|F'_2|\,\ee^{\ii\delta^{(-)}}\right\},\cr
F_{S,K}=&\,q_2^{-1/2}\left\{(\cos\theta)|F'_2|\,\ee^{\ii\delta^{(-)}}- 
(\sin\theta)|F'_1|\,\ee^{\ii\delta^{(+)}}\right\}.\cr}
\equn{(2.6b)}$$
Near $\bar{K}K$ threshold or for small inelasticity, 
$\delta_S\simeq\delta^{(+)}\simeq\phi_\pi$.

\booksection{3. The  partial wave amplitudes from the experiment of Hyams et al.}

\noindent
We will here follow DGL and ACCGL and take 
the partial wave amplitudes as measured by Hyams et al.,\ref{4} although
 later we will also discuss other  sets of $\pi\pi$ scattering data, as well as  
data\ref{5} on $\pi\pi\to\bar{K}K$.
Hyams et alii give three representations for their data: an energy-independent 
phase shift analysis that yields the values of the phase shift $\delta_\pi(s)$,
 and of the elasticity parameter 
$\eta(s)$, from 
$\pi\pi$ threshold to $s^{1/2}\simeq1.9\,\gev$; 
an energy-dependent
 parametrization of the K-matrix that interpolates these data in the whole range; 
and a second parametrization with a constant K-matrix 
that represents the data around $\bar{K}K$ threshold.

For the second,  Hyams et alii write  [Eq.~(12a) and Table~1 in ref.~4]
$$K_{ab}(s)=\alpha_a\alpha_b/(s_1-s)+\beta_a\beta_b/(s_2-s)+\gamma_{ab},
\equn{(3.1)}$$
where
$$\eqalign{
s^{1/2}_1=&\,0.11\pm0.15,\quad s^{1/2}_2=1.19\pm0.01;\cr
\alpha_1=&\,2.28\pm0.08,\quad\alpha_2=2.02\pm0.11;\quad
 \beta_1=-1.00\pm0.03,\quad\beta_2=0.47\pm0.05;\cr
\gamma_{11}=&\,2.86\pm0.15,\quad \gamma_{12}=1.85\pm0.18,\quad
 \gamma_{22}=1.00\pm0.53.}
\equn{(3.2)}$$
The numbers here are in the appropriate powers of \gev.

In the energy range around $\bar{K}K$ threshold,  $0.9\,\gev\leq s^{1/2}\leq1.1\,\gev$, 
Hyams et alii (p.~148 of ref.~4)
 find that their data may be represented by a  constant
K-matrix with
$$K_{11}=1.0\pm0.4\;{\gev}^{-1},\quad K_{12}=-4.4\pm0.3\;{\gev}^{-1},
\quad K_{22}=-3.7\pm0.4\;{\gev}^{-1}.
\equn{(3.3)}$$
The sign of $K_{12}$ is undefined. We have chosen in 
(3.3) a sign opposite to that of Hyams et al.,\ref{4} to agree with 
what the same authors get from the 
energy-dependent K-matrix; see below, Eq.~(3.4).
This is  somewhat different from what (3.2) gives at $\bar{K}K$ threshold:
evaluating $K(s)$ with the central values in (3.2) one finds 
$$K_{11}(4m^2_K)=-0.17\;{\gev}^{-1},\quad K_{12}(4m^2_K)=-4.0\;{\gev}^{-1},
\quad K_{22}(4m^2_K)=-2.7\;{\gev}^{-1}.
\equn{(3.4)}$$

Before starting with the actual analyses it is perhaps convenient to remark that 
what follows from {\sl experiment} is the {\sl energy-independent} set of 
phase shifts and elasticity parameters. 
The energy-dependent representations are model dependent. This is 
particularly true of (3.1),  where one makes the choice of a specific functional form; 
the results vary somewhat if using other parametrizations.

\booksection{4. The phase $\phi_\pi$}

\noindent
We will here consider the value of the phase $\phi_\pi(s)$ that follows from 
the experimental analysis given above. 
Although  $\phi_\pi(s)$ is different from the quantities $\delta_S(s)$ and $\delta^{(+)}(s)$,
 which are the ones that intervene in the evaluation of the scalar form factor, 
they follow the same pattern. This was noted by ACCGL, who discuss 
$\phi_\pi$ in detail to illustrate their conclusions on 
$\delta_S$, and, indeed, it 
can be verified without too much trouble with the formulas of 
\sect~2: explicitely for $\delta^{(+)}$ and to first order in $q_2$ or in $\epsilon$ for $\delta_S$ 
(the exact result for the last requires solving two coupled integral equations). 

The advantage of $\phi_\pi$ is that it is given by the 
simple equation (2.4) in terms of the 
observable quantities $\delta_\pi$, $\eta$. 
This will allow us to simplify the discussion enormously; 
in particular, it will let us use simple parametrizations 
of $\delta_\pi$, $\eta$ above $\bar{K}K$ threshold, which is the region where 
there is disagreement between the evaluation of ref.~2 and DGL, ACCGL. 
This simplification is unnecessary in the sense that the results are almost identical 
to what one finds with the full K-matrix (that we will present later); 
but it permits us to exhibit, with great clarity, both the mechanisms at work and 
the issues involved.

To calculate $\phi_\pi$ around and above $\bar{K}K$ threshold we take
$$\eqalign{
\delta_\pi(s)=&\,\pi+d(s)+\dfrac{q_2}{s^{1/2}}c(s),\cr
\eta(s)=&\,1-\epsilon(s)\cr
}\equn{(4.1)}$$   
and approximate, for $0.95\,\gev\lsim s^{1/2}\lsim1.35\,\gev$,
$$\eqalign{
d(s)=&\,d_0={\rm constant},\quad c(s)=c_0={\rm constant},\cr
\epsilon(s)&\,=\left(\epsilon_1\dfrac{q_2}{s^{1/2}}+
\epsilon_2\dfrac{q_2^2}{s}\right)\dfrac{M^2-s}{s},\quad M=1.5\;\gev.
}\equn{(4.2)}$$
In the region immediately below $\bar{K}K$ threshold we replace $q_2$ by $|q_2|$ 
in (4.1). 

The energy-independent set of data in ref.~4 are well fitted
 with the numbers\fnote{We have actually followed the fit of ref.~6, which 
takes into account other data sets 
and is slightly below, both for $\delta_\pi$ and $\epsilon$,
from what Hyams et al. give, at the upper energy range.}
$$c_0=5\pm1,\quad \epsilon_1=6.4\pm0.5,\quad\epsilon_2=-16.8\pm1.6
\equn{(4.3)}$$ 
and we will leave the value of $d_0$ (which is small) free for the moment. 
It will turn out that a key quantity in the analysis is the phase 
shift at $\bar{K}K$ threshold, 
$\delta_\pi(4m^2_K)$, and we want to be able to vary this.

\booksubsection{4.1. The  phase $\phi_\pi$ of DGL, ACCGL}

\noindent
The authors of refs.~1,~3 take the K-matrix of Hyams et al.,\ref{4}
 with the central values as given in 
(3.2). 
What is important for us here is that this implies that the central value 
of $\delta_\pi(4m^2_K)$ is less than 180\degrees:
$$\delta_\pi(4m^2_K)=175\degrees.
\equn{(4.4a)}$$
To reproduce this, we have to take $d_0$ in (4.2) negative and equal to
$$d_0=-0.087.
\equn{(4.4b)}$$
Care has to be exercised when crossing the energy $s_0$ at which 
$\delta_\pi(s)$ equals $\pi$, which, with (4.3) and (4.4b), occurs at 
$s_0^{1/2}=992.6~\mev$, 
$$\delta_\pi(s_0=(992.6\;\mev)^2)=\pi,$$
and where (2.4) is singular.
For the moment, we will tackle this by starting below $s_0$ and requiring continuity 
of {\sl the phase} $\phi_\pi(s)$ across $s_0$. 
This we will call a solution of {\sl Type~I}, and is like what ACCGL find; 
indeed, the corresponding values of $\delta_\pi(s)$, $\phi_\pi(s)$,  shown in Fig.~1, 
are practically identical to those in the Fig.~1 in ACCGL
in the relevant region, around and above $\bar{K}K$ threshold. 
As can be seen in both figures, in the region $s^{1/2}\sim1.35\,\gev$, where inelasticity 
is negligible, 
$\delta_\pi(s)$ and $\phi_\pi(s)$ differ by $\pi$. 
 $\delta_S(s)$ and $\delta^{(+)}(s)$ are  very 
similar to $\phi_\pi(s)$ and thus also differ by $\pi$ from $\delta_\pi(s)$.

\topinsert{
\setbox0=\vbox{\hsize16.2truecm{\epsfxsize 14.6truecm\epsfbox{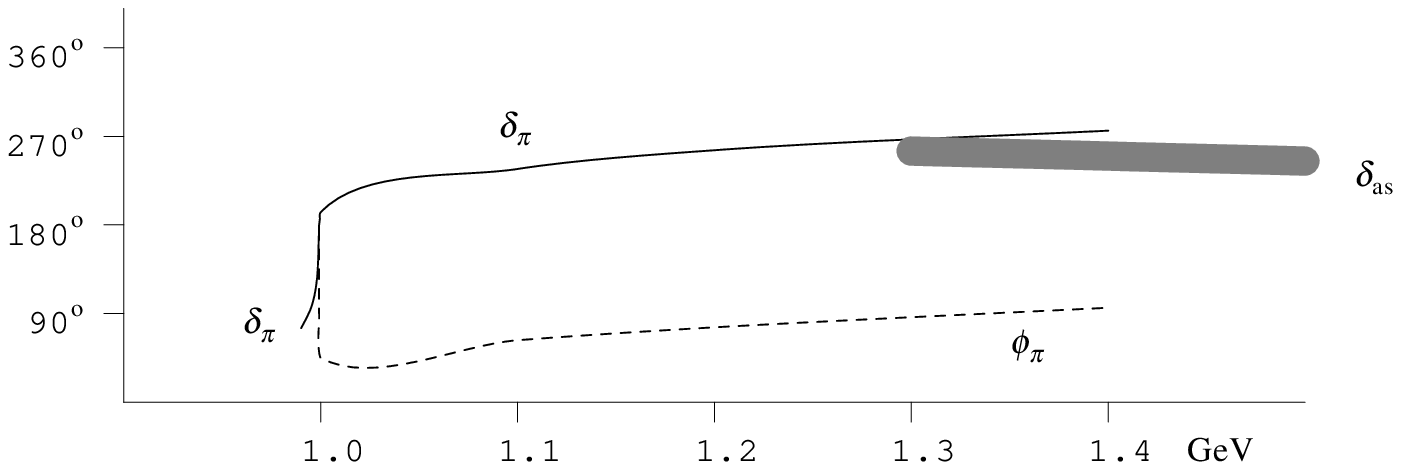}}} 
\setbox6=\vbox{\hsize 12truecm\captiontype\figurasc{Figure 1. }{
The phases $\delta_\pi(s)$ (continuous line) 
and $\phi_\pi(s)$ (dashed line) corresponding to a
Type~I solution. Note that, below 
$\bar{K}K$ threshold 
$\delta_\pi(s)=\phi_\pi(s)$, hence $\phi_\pi(s)$ shows 
a very pronounced spike at $s=4m^2_K$. 
The asymptotic phase (to be defined below) $\delta_{\rm as.}$
 is represented 
by the thick gray line.}\hb} 
\centerline{\tightboxit{\box0}}
\bigskip
\centerline{\box6}
}\endinsert

{\sl This is the key remark of ACCGL\/}: the phases $\delta_\pi(s)$ and $\phi_\pi(s)$, $\delta_S(s)$ 
are {\sl not} equal above $s^{1/2}\sim1.1\,\gev$, 
but rather one has
$$\delta_S(s)\simeq\delta^{(+)}(s)
\simeq\phi_\pi(s)\simeq\delta_\pi(s)-\pi,\quad s^{1/2}\gsim1.1\,\gev.
$$
This  
 accounts for the difference between the results of refs.~1,~3 (DGL,~ACCGL)
 and my previous results\ref{2} for the 
integral (1.7), hence for the different values of the scalar radius.

\topinsert{
\setbox0=\vbox{{\psfig{figure=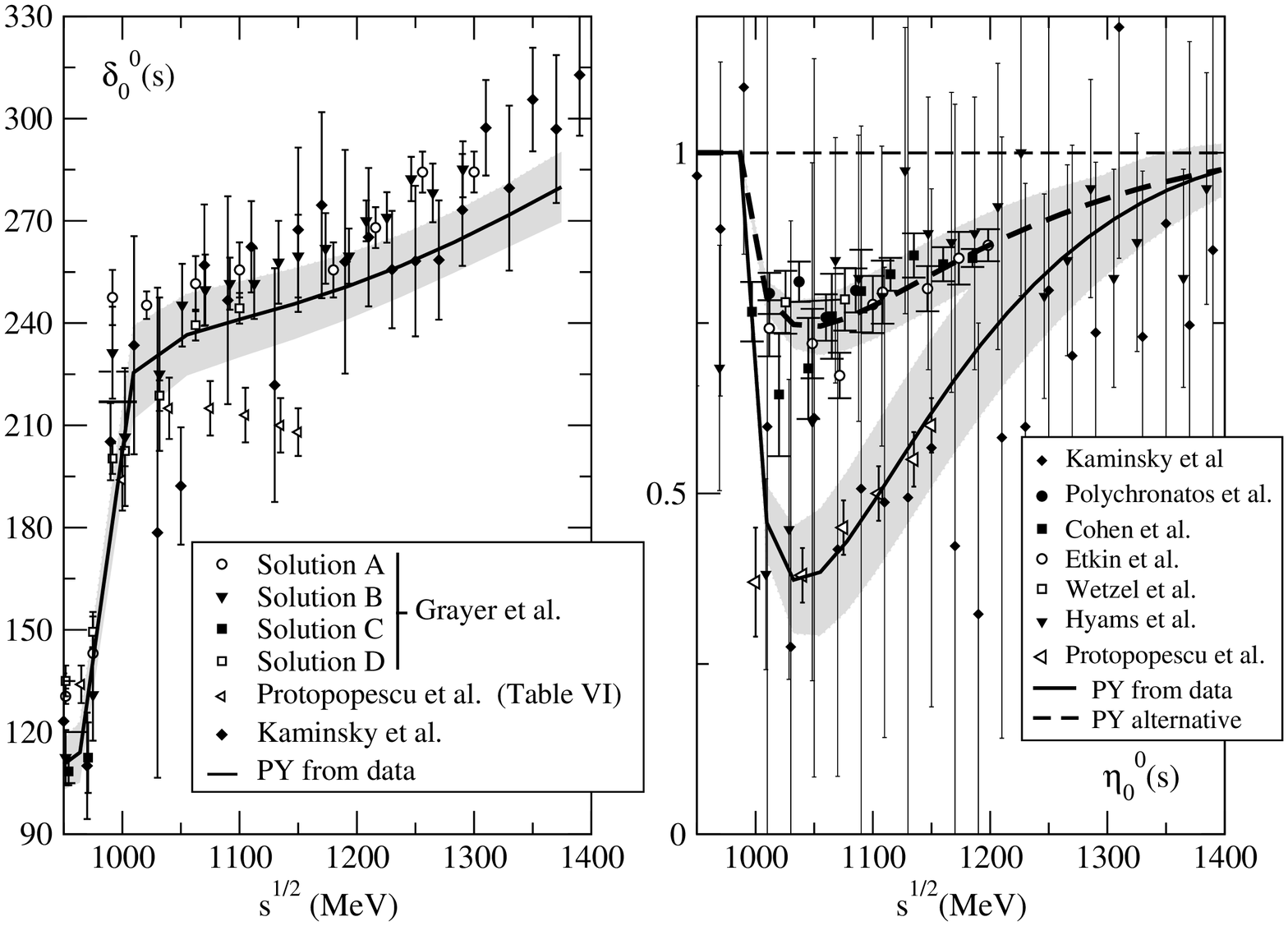,width=13truecm,angle=-90}}} 
\setbox6=\vbox{\hsize 12.5truecm\captiontype\figurasc{Figure 2. }{Fit to the
 $I=0$, $S$-wave  inelasticity
 and phase shift  between 
 950 and 1400 \mev, from ref.~6 [so that the 
formula used for $\delta_0^{(0)}\equiv \delta_\pi$ 
is slightly different from (4.1)], and  data  from refs.~4,~5,~8 and 11.
The shaded bands correspond to $1\,\sigma$ variation in 
the parameters of the fits.
The fit to the phase shift corresponds to  $d_0=0$.
The 
difference between the determinations of $\eta$ from $\pi\pi\to\pi\pi$ 
({\sl PY from data} in the figure) and 
from $\pi\pi\to\bar{K}K$ ({\sl PY alternative}) is apparent here.
}\hb} 
\centerline{\tightboxit{\box0}}
\bigskip
\centerline{\box6}
\medskip
}\endinsert

The situation, however, is not as simple as ACCGL seem to believe.
First of all, the {\sl inelasticity} given in ref.~4 
is much overestimated. 
After that paper was written, a number of experiments have appeared\ref{5} 
in which the cross section $\pi\pi\to\bar{K}K$ was measured. 
Since there are no isospin-2 waves in $\pi\pi\to\bar{K}K$ scattering, 
and the  $\pi\pi-\bar{K}K$ coupling is very weak for P, D waves,
 it follows 
that measurements of the differential cross section for 
 $\pi\pi\to\bar{K}K$ give directly $1-\eta^2$ with good accuracy.
On the other hand, $\pi\pi$ scattering experiments like those of 
 Hyams et al.\ref{4} only 
measure the $\pi\pi\to\pi\pi$ cross section, so that $\eta$ is  obtained 
less precisely here: not only 
the $\pi\pi$ cross section depends on both $\delta_\pi$, $\eta$, 
but other waves (notably S2, P and D0) interfere. 
Thus, these more recent, $\pi\pi\to\bar{K}K$ based, experimental values\ref{5}
 for $\eta$ are much more reliable than the older ones, in particular than those of ref.~4.

The value of the inelasticity the experiments in ref.~5 give is 
 about a {\sl third} of 
what (4.3) indicates: $\eta$
 can be fitted with\ref{6}
$$\epsilon_1=2.4\pm0.2,\quad \epsilon_2=-5.5\pm0.8.
\equn{(4.5)}$$
The difference is shown graphically in Fig.~2.

If we now use (4.5) instead of (4.3) to calculate $\phi_\pi(s)$, 
keeping $\delta_\pi(s)$ fixed, a surprising result occurs: 
$\phi_\pi(s)$ does {\sl not} become closer to $\delta_\pi(s)$  above 
the point $s_0$; on the contrary, it  moves  
closer to $\delta_\pi-\pi$. 
In fact, one can decrease the inelasticity to zero, $\epsilon(s)\to0$, 
{\sl keeping $\delta_\pi(s)$ fixed}, and one finds that 
$$\eqalign{
\phi_\pi(s)\to\delta_\pi(s),\quad s^{1/2}<s^{1/2}_0=992.6\;\gev;\cr
\phi_\pi(s)\to\delta_\pi(s)-\pi,\quad s^{1/2}>s^{1/2}_0=992.6\;\gev.\cr
}
\equn{(4.6)}$$
That is to say: contrary to 
physical expectations, the limit of zero inelasticity does not coincide 
with inelasticity zero for, if we set $\epsilon(s)\equiv0$, then 
 $\delta_\pi(s)$ and $\phi_\pi(s)$ should be identical. 
This phenomenon was noticed by ACCGL who, however, failed to attach to it 
the due importance.
As a matter of  fact, the situation is even more complicated, as will be shown below: 
if we leave $\eta$ fixed but vary $d_0$ in (4.2) across zero to a positive number, 
however small, the resulting $\phi_\pi$
 is not continuous when $d_0$ crosses zero: it jumps by $\pi$.

What is the reason for this peculiar behaviour of $\phi_\pi$?
It is not difficult to identify: Eq.~(2.4) does {\sl not} determine $\phi_\pi$, but only 
its tangent. 
Thus, $\phi_\pi$ is only fixed up to a multiple $N\pi$. 
$N$ may be set to zero below the point $s_0$ where $\delta_\pi(s)$ crosses 
$\pi$, by requiring that $\phi_\pi(4m^2_K)=\delta_\pi(4m^2_K)$ and continuity 
above this. 
However, Eq.~(2.4) shows that $\tan\phi_\pi(s)$
 is {\sl discontinuous} when $s$ crosses $s_0$. Therefore, we may well add $\pi$ to 
the $\phi_\pi(s)$ of the Type~I solution found above, 
in the region  $s>s_0$, since this does not change its tangent. 
We then find what we call a solution of {\sl Type~Id} ({\sl d} for discontinuous), 
depicted in Fig.~3. 
In this case, $\phi_\pi(s)$ is {\sl not} continuous across $s_0$, but does tend\fnote{To
 be precise, one should remark that
this limit applies {\sl in the mean}; the isolated point 
$s_0$ remains singular. Convergence in the mean, however, is sufficient to ensure 
 convergence of integrals involving $\phi_\pi$.} to
 $\delta_\pi(s)$, for all values of $s$, when the inelasticity tends to zero. 
It is also continuous (in the mean) for $d_0$ around zero. 
ACCGL appear to be  unaware of the existence of solutions of Type~Id.

\topinsert{
\setbox0=\vbox{\hsize16.2truecm{\epsfxsize 14.6truecm\epsfbox{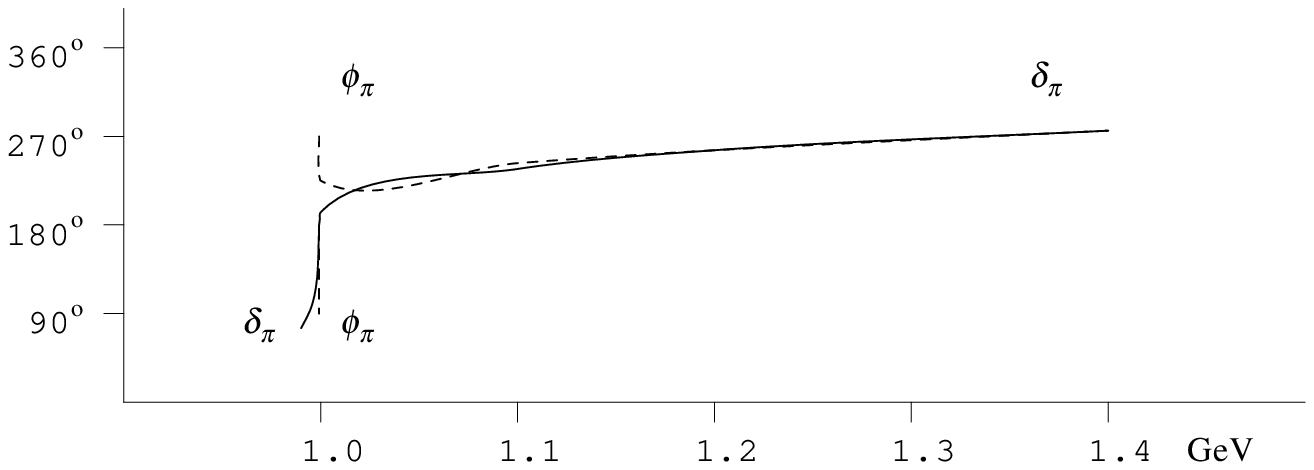}}} 
\setbox6=\vbox{\hsize 12truecm\captiontype\figurasc{Figure 3. }{
The phases $\delta_\pi(s)$ (continuous line) 
and $\phi_\pi(s)$ (dashed line) corresponding to a solution of
Type~Id. The spike that appeared in \fig~1 is now accompanied by 
a jump of $\phi_\pi(s)$, at $s=s_0$.}\hb} 
\centerline{\tightboxit{\box0}}
\bigskip
\centerline{\box6}
}\endinsert

It is not clear which of the two solutions, Type~I or Type~Id, should be considered 
correct: both types look awry. 
In fact, we will show that both  Type~I and Type~Id are, with all probability,
 spureous solutions, 
 artifacts due to the use of the parametrization (3.1), (3.2)
over too wide a range, and with too little experimental 
information.

\goodbreak
\booksubsection{4.2. The correct $\phi_\pi$}

\noindent
We next repeat the 
calculations of the previous section, 
but we will now assume that $\delta_\pi(4m^2_K)$ is {\sl larger} than $\pi$, 
so that $d_0$ is {\sl positive}. 
To get this it is  sufficient to alter a little the parameters in 
(3.2). For example, if we move only one parameter by $1\,\sigma$, just  replacing 
in (3.2)
$$\alpha_1\to2.20=2.28-0.08,
\equn{(4.7)}$$
then  $\delta_\pi(4m^2_K)$ becomes 185\degrees. 
Note that 
$\delta_\pi(s)$ is almost unchanged by this, as may be seen by comparing Figs.~1 and 4.
 The only important effect of the change in (4.7) 
is to push $\delta_\pi(4m^2_K)$ from a bit below to a bit above 180\degrees; 
but then,  this is a key point,  as we will see.

A value for 
$\delta_\pi(4m^2_K)$ above 180\degrees\ follows also for 
$s_1=0$, $\gamma_{11}=3.0$ (as in ref.~7), which values are both less than $1\,\sigma$ off the 
central values in (3.2). 
In fact, a value 
 $\delta_\pi(4m^2_K)> 180\degrees$ can already be obtained  with
only  a $\tfrac{1}{2}\,\sigma$ change, 
$$\alpha_1\to2.24=2.28-0.04.$$ 
Thus, a value  $\delta_\pi(4m^2_K)> 180\degrees$ is perfectly 
compatible with the energy-dependent parametrization of 
Hyams et alii, Eqs.~(3.1), (3.2), when errors are taken into account.

\topinsert{
\setbox0=\vbox{\hsize16.2truecm{\epsfxsize 14.6truecm\epsfbox{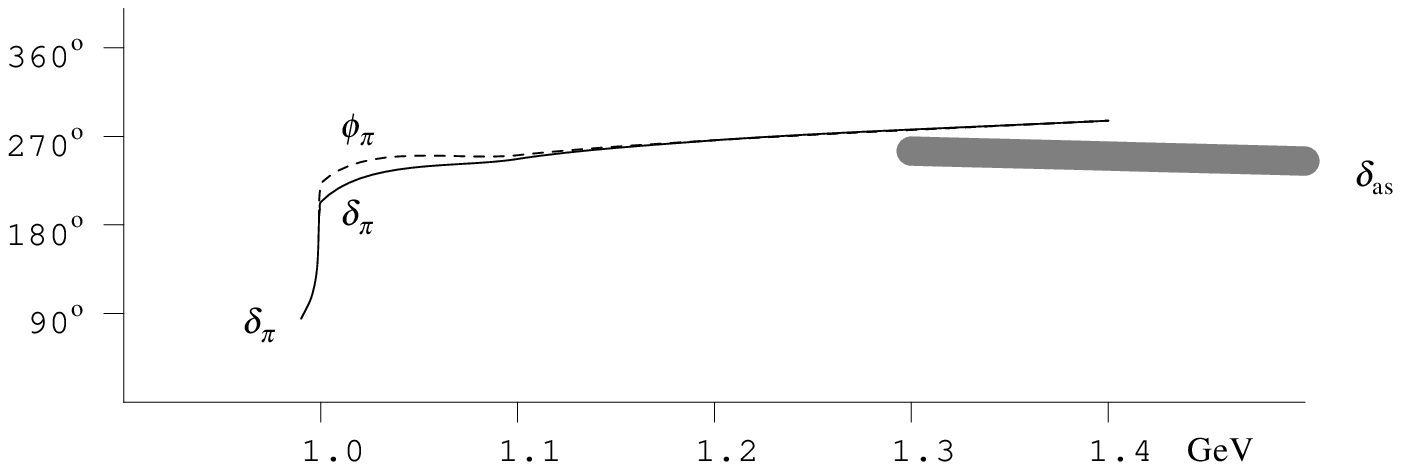}}} 
\setbox6=\vbox{\hsize 12truecm\captiontype\figurasc{Figure 4. }{
The phases $\delta_\pi(s)$ (continuous line) 
and $\phi_\pi(s)$ (dashed line) corresponding to the solution of
Type~II.   As in Fig.~1, the thick gray line is the asymptotic phase $\delta_{\rm as.}$.}\hb} 
\centerline{\tightboxit{\box0}}
\bigskip
\centerline{\box6}
}\endinsert

We will use (4.7) for simplicity in the discussion and 
will thus repeat the calculations with
$$\delta_\pi(4m^2_K)=185\degrees,\quad d_0=+0.087.
\equn{(4.8)}$$
In the present case, and as is obvious from (2.4), $\phi_\pi(s)$ 
is never singular and it  stays {\sl above} $\delta_\pi(s)$, 
up to the energy $s^{1/2}\sim1.3\,\gev$ where 
$\delta_\pi(s)$ crosses $3\pi/2$, remaining close to it afterwards.\fnote{In fact, 
 over the whole range, the difference between $\phi_\pi$ and $\delta_\pi$ is smaller than 
the experimental errors of the last: compare Figs.~2 and 4.}

{\sl This property is actually quite general}, not tied to the specific approximations 
(4.2),~(4.8), and depends only on the fact that 
 $\delta_\pi(s)$ is an increasing function of $s$ and that $\delta_\pi(4m^2_K)>\pi$.
This is all 
we need for 
$\phi_\pi$. To get the analogous property for $\delta^{(+)}$, $\delta_S$ 
we also require that $\det {\bf K}(4m^2_K)<0$, something that is amply satisfied with the 
parameters of (3.2), (3.3) or, more generally, if, 
as implied by SU(3) ch.p.t., one has $\tan\delta_K<0$ near $\bar{K}K$ 
threshold  (see below).

\topinsert{
\setbox0=\vbox{\hsize11.8truecm{\epsfxsize 10.2truecm\epsfbox{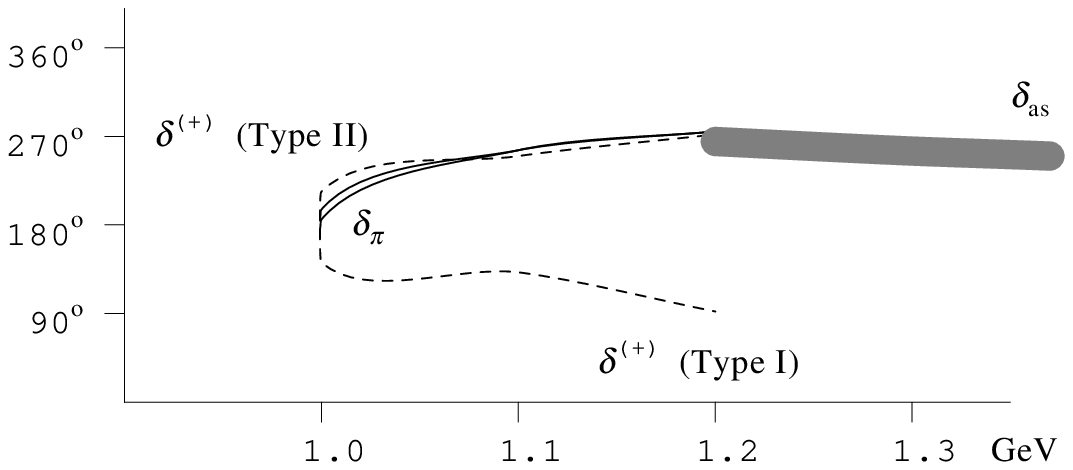}}\hfil} 
\setbox6=\vbox{\hsize 4truecm\captiontype\figurasc{Figure 5. }{
The phases $\delta_\pi(s)$ (continuous lines) 
and $\delta^{(+)}(s)$ (dashed lines)  evaluated in the 
K-matrix formalism, Eq.~(3.1), with the central values of the 
parameters given in
(3.2) (Type~I) or (4.7), Type~II.
[The two lines for $\delta_\pi$ correspond also 
to (3.2), (4.7)]. 
 The asymptotic phase $\delta_{\rm as.}$ (thick gray line)
is also shown.}\hb} 
\line{{\tightboxit{\box0}}\hfil\box6}
\bigskip
}\endinsert

A set of phases with these properties we will call a solution of  {\sl Type~II}.
In the specific case (4.8) we find the $\delta_\pi(s)$, $\phi_\pi(s)$ depicted in Fig~4.
Note that  $\delta_\pi(s)$ and $\phi_\pi(s)$ are near each other all the time, 
as one expects physically since the inelasticity is small; 
this is particularly important in view of the results of refs.~5. 
Unlike what happened in solutions of Type~I, or Type~Id, the phase 
$\phi_\pi(s)$ is now a smooth function
 both of $\epsilon(s)$ and of $s$.

These results are not new. 
They were amply discussed more than thirty years ago, in connection 
with the eigenphases $\delta^{(\pm)}(s)$, by the present author in ref.~7. 
There it was noted that, by going from the  values of the K-matrix parameters in (3.2) to 
values like those in 
(4.7), the eigenphase $\delta^{(+)}(s)$ changes from a fast
decrease above the $\bar{K}K$ threshold, diverging  from $\delta_\pi(s)$ by $\sim\pi$ 
(as does $\phi_\pi$ in a solution of Type~I, see \fig~1), to 
increasing above  $\bar{K}K$ threshold with increasing $s$,
  staying close, but a bit above, 
$\delta_\pi(s)$ (again, as does $\phi_\pi$ in a solution of  Type~II, \fig~4). 
The reader may compare our Figs.~1,~4 here with 
Fig.~2 in ref.~7. 
In  ref.~7 the M-matrix (${\bf M}={\bf K}^{-1}$) 
parametrization of experimental data of 
Protopopescu et al.\ref{8} is also considered, 
and the same phenomenon is observed  (Fig.~1 in ref.~7).

We give in \fig~5
 the eigenphases $\delta^{(+)}$ 
corresponding to Type~I and Type~II solutions.
Here $\delta_\pi$, as well as the eigenphase $\delta^{(+)}$, 
are evaluated with the K-matrix formalism, Eq.~(3.1). 
For Type~I we took the parameters (3.2); for Type~II, those in (4.7).  
Our Fig.~5 here agrees with the corresponding parts of Figs.~1,~2 in ref.~7.

\booksection{5. The value of $\delta_\pi(4m^2_K)$ }

\noindent
As is obvious from the previous discussion, 
a key quantity in this analysis is the $\pi\pi$ phase at
$\bar{K}K$ threshold, $\delta_\pi(4m^2_K)$. 
If this is smaller than $\pi$, we have a  situation of Type~I; 
if, on the contrary, $\delta_\pi(4m^2_K)>\pi$, 
we have a solution of Type~II and, in particular, we can approximate
$$\delta_S(s)\simeq\delta^{(+)}(s)\simeq\phi_\pi(s)\simeq\delta_\pi(s),
$$
as was done in ref.~2. 

It should be clear that the 
parametrization (3.1), (3.2) is not a good guide to find the value of
 $\delta_\pi(4m^2_K)$. Not only  $\delta_\pi(4m^2_K)$ crosses 180\degrees\ 
when varying the parameters in (3.2) within 
their errors (as we have shown before) but, more to the point,                          
(3.1) was devised to furnish an {\sl approximate} 
representation of $\delta_\pi(s)$, $\eta(s)$ in the 
whole range $4M^2_\pi$ to 
$1.9^2\,\gev^2$. This may easily create local distortions; and, 
in fact, such distortions are expected. The inelasticity of 
ref.~4 is overestimated, as proven by the more 
precise measurements of ref.~5:  this will influence   
the phase $\delta_\pi$ above 1~\gev, hence, via the 
parametrization, around $\bar{K}K$ threshold. Such a distortion 
also occurs in the evaluation of Au et al.,\ref{9} who make a fit to
 $\eta$ and $\delta_\pi$, based on data of ref.~4, over the whole energy range, 
 which fit leads to a value of $\delta_\pi(4m^2_K)$ smaller than 180\degrees: 
see Fig.~4 in ref.~9.
 We certainly need something more precise in the 
vicinity of the $\bar{K}K$ threshold, since $\delta_\pi(4m^2_K)$ is so near $\pi$.

For this we have several
 possibilities: the constant K-matrix fit around $\bar{K}K$ 
threshold of Hyams et al.;\ref{4} 
the energy-independent analysis of this same reference;  
 the results of other experiments; or certain theoretical arguments. 
As for the first, if we take the  values $K_{ab}$ in (3.3), obtained from a fit 
to data from 0.9~\gev\ to 1.1~\gev, we find
$$\delta_\pi(4m^2_K)=205\pm8\degrees,
\equn{(5.1)}$$
$3\,\sigma$ above 180\degrees. A value above 180\degrees\ is, of course, also found if 
interpolating the 
energy-independent analysis of  ref.~4. 
The data of Protopopescu et al.\ref{8} are not sufficiently precise to 
discriminate  whether $\delta_\pi(4m^2_K)$ is below or above 180\degrees:
 for some of the solutions in ref.~8, $\delta_\pi(4m^2_K)$ 
is below, and for others above 180\degrees, but in all cases, the errors 
cover the value  180\degrees\ (at any rate, the elasticity parameter of 
Protopopescu et al. is  incompatible with 
$\pi\pi\to\bar{K}K$ results).
However, a value clearly  above  180\degrees\ is found 
if extrapolating downward the experimental results of ref.~10  
(the phase shift is only measured for $s^{1/2}>1\,\gev$). 
This gives\fnote{For the favoured solution 
in ref.~10 which, incidentally, is the one 
with values of $\eta(s)$ more compatible with measurements based on $\pi\pi\to\bar{K}K$.
 For other solutions $\delta_\pi(4m^2_K)$ is even larger, except for 
one that yields a value near $180\degrees$. This last one, hovever, has an 
 elasticity parameter incompatible with 
$\pi\pi\to\bar{K}K$ results.}
 $\delta_\pi(4m^2_K)=203\pm7\degrees$, including estimated systematic errors. 
A value 
 $\delta_\pi(4m^2_K)>180\degrees$ is also found   
in all five solutions of Grayer et al.:\ref{11} cf.~\fig~31 there. 
Finally, Kami\'nski et al.\ref{11} find $\delta_\pi(4m^2_K)=190\pm25\degrees$. 
The experimental information thus very strongly favours a value  
 $\delta_\pi(4m^2_K)>180\degrees$, and hence a solution of Type~II.

There are two other independent, theoretical arguments in favour of  a solution of Type~II. 
The first
is based on chiral SU(3) calculations: unitarized SU(3) ch.p.t. 
 produces central values of $\delta_\pi(4m^2_K)$ above 190\degrees\ (with 
a value around 200\degrees\ favoured; see 
for example ref.~12). Moreover, in 
Type~II solutions, with the parameters in (4.7), one has 
a real part of the 
$\bar{K}K$ scattering length $a_r(\bar{K}K)\simeq-0.46\,M_{\pi}^{-1}$, 
in   agreement with the unitarized current algebra (ch.p.t.) result 
that gives $a_r(\bar{K}K)\simeq-0.5\,M_{\pi}^{-1}$.  

The second, more serious indication, is that 
the phase $\delta_S(s)\simeq\phi_\pi(s)$  for Type~II solutions joins smoothly 
the result furnished by the perturbative QCD evaluation of $\delta_S(s)$, 
while a Type~I solution $\delta_S(s)$ lies clearly below. 
We now turn to this.

\booksection{6. The phase $\delta_S(t)$ at large $t$ from QCD} 

\noindent
Using the evaluations in ref.~13 it is easy to 
get that, to leading order in 
the QCD coupling $\alpha_s$, one has
$$F_{S,\pi}(t)=\dfrac{4\pi[m^2_u(\nu^2)+m^2_d(\nu^2)]C_F\alpha_s(\nu^2)}{-3t}I,
\equn{(6.1)}$$
where, neglecting quark and pion masses, 
$$I=\tfrac{1}{2}\Bigg\{\int_0^1\dd\xi\,\dfrac{\psiv^*(\xi,\nu^2)}{1-\xi}
\int_0^1\dd\eta\,\dfrac{\psiv(\eta,\nu^2)}{(1-\eta)^2}+
\int_0^1\dd\xi\,\dfrac{\psiv^*(\xi,\nu^2)}{(1-\xi)^2}
\int_0^1\dd\eta\,\dfrac{\psiv(\eta,\nu^2)}{1-\eta}\Bigg\}.
\equn{(6.2)}$$
Here $\nu^2$ is the renormalization point and $\psiv$ is 
the partonic wave function of the pion, defined by
$$(2\pi)^{3/2}\langle0|{\cal S}:\bar{d}(0)\gamma_\lambda\gamma_5D_{\mu_1}\cdots D_{\mu_n}
u(0):|\pi(p)\rangle=\ii^{n+1}p_\lambda p_{\mu_1}\cdots p_{\mu_n}\int_0^1\dd\xi\,\xi^n\psiv(\xi,\nu^2).
$$
The $D_\mu$ are covariant derivatives, and $\cal S$ means symmetrization. 
The function $\psiv$ is the same that appears in 
the evaluation of the vector form factor, and thus\ref{13}
$$\psiv(\xi,\nu^2)\simeqsub_{\nu\to\infty}\xi(1-\xi)6\sqrt{2}\,f_\pi.
\equn{(6.3)}$$
If we input (6.3) into 
(6.2) we get a divergent result. 
This divergence may be traced to the fact that we have 
neglected quark and pion masses, and may be cured by defining the form factor 
not for external momenta $p^2=p'^2=0$, but with  $p^2=p'^2=t_0$, 
$t_0$ being a fixed number; for example, we could take $t_0=M^2_\pi$. 
Then we choose $\nu^2=-t$ (for spacelike $t$) and find the asymptotic 
behaviour
$$F_{S,\pi}(t)\simeqsub_{t\to\infty}
\dfrac{48\pi[\,m^2_u(-t)+m^2_d(-t)]C_Ff^2_\pi\alpha_s(-t)\log(-t/t_0)}{-t}
\to \dfrac{C[\,m^2_u(-t)+m^2_d(-t)]f^2_\pi}{-t}
\equn{(6.4)}$$
with $C=576\pi^2C_F/(33-2n_f)$,  and $n_f$ is the number of quark flavours, 
that we take equal to three.

Unfortunately, the value of the constant $C$ is changed
 when higher order corrections are included. 
These  have the same structure as (6.4),  with higher powers
$[\,\alpha_s(-t)\log(-t/t_0)]^n$ which are not suppressed at large $t$. 
Therefore, the constant $C$ gets contributions from all orders of perturbation 
theory 
with the result that the final value is unknown.
However, it is very likely that the structure $[({\rm Constant})\times\sum{m}^2_i(-t)/t]$ remains. 
This is sufficient to get 
a prediction for the asymptotic phase:
$$\delta_S(s)\simeqsub_{s\to\infty}\delta_{\rm as.}(s)=
\pi\left\{1+\dfrac{2d_m}{\log (s/\lambdav^2)}\right\},\quad d_m=12/(33-2n_f).
\equn{(6.5)}$$
Here $\lambdav$ is the QCD parameter; in our calculations here we 
have allowed it to  vary in  
the range  
  $0.1\,{\rm GeV}^2\leq\lambdav^2\leq0.35\,{\rm GeV}^2$.
$\delta_{\rm as.}(s)$ is the phase plotted in Figs.~1,~4,~5 where it 
is seen very clearly that it is consistent with Type~II solutions, but 
not with the Type~I solution of ACCGL.

\booksection{7. Conclusions} 

\noindent
There are other methods for finding 
 directly $\bar{l}_4$, of which we only mention two. 
One can evaluate on the lattice the  dependence of the quark 
condensate on the quark masses;\ref{14} or one can fit  $\bar{l}_4$
to the experimental $\pi\pi$ scattering lengths and effective range parameters obtained 
from  experimental data,\ref{6} using ch.p.t. to one loop.\ref{15}
The results are summarized below, where we also repeat the results of refs.~1,~2,~3:

$$\bar{l}_4=\cases{
4.4\pm0.2\quad \hbox{[refs.~1, 3]}\cr
5.4\pm0.5\quad \hbox{[ref.~2]}\cr
4.0\pm0.6\quad \hbox{[lattice calculation, ref.~14]}\cr
7.2\pm0.7\quad \hbox{[fitting $a_l^{(I)}$, $b_0^{(I)}$, $b_1$, ref.~15]}.\cr }
\equn{(7.1)}$$
This is inconclusive; lattice calculations are known to suffer from large systematic errors, and the 
number following from the fit to experimental data is affected by higher order corrections, 
which the evaluation in ref.~15 does not take into account.
We have to fall back on our previous discussion, involving the phase of the scalar form factor.

In this case, and as we have shown, we have two types of solution: 
Type~I, that occurs
 when $\delta_\pi(4m^2_K)<\pi$, and Type~II, when $\delta_\pi(4m^2_K)>\pi$. 
The correctness of a  solution of Type~I, 
which is the one used in the evaluations of 
DGL, ACCGL
 is very unlikely: 
the experimental indications\ref{4,10,11} favour 
values $\delta_\pi(4m^2_K)>\pi$. 
Moreover, in Type~I solutions one has a discontinuous phase $\phi_\pi$, 
 when the inelasticity tends to 
zero. Type~I solutions  also exhibit a phase $\phi_\pi$ which is not continuous when 
$\delta_\pi(4m^2_K)$ moves around $\pi$. 
Finally, Type~I solutions give a phase $\delta_S(s)$ rather different from 
what perturbative QCD suggests, Eq.~(6.5), at large $s$. 
We think that Type~I solutions are spureous, unphysical solutions, 
which appear only because one tries to fit, with too  
 simple a formula, and without enough experimental 
information, the whole energy range from $\pi\pi$ threshold to 1.9~\gev, 
which distorts the results in the region of $\bar{K}K$ threshold.
This last conjecture is confirmed by the 
evaluations of Moussallam.\ref{16}
This author uses, like DGL, ACCGL, fits that represent
 the quantities $\delta_\pi$ and $\eta$ 
over the whole energy range; in particular, the fit of Au et al.\ref{9}
Such parametrization gives $\delta_\pi(4m^2_K)\simeq173\degrees$, 
hence a Type~I solution and thus, not surprisingly,  Moussallam 
finds a value for 
$\langle r^2_S\rangle$ similar to that of DGL.

Although this is not very important, because the very starting point of DGL, ACCGL 
(a Type~I solution) is unlikely to be correct, one may question 
 the methods of error analysis of these authors.
 As we discussed above, 
a value  $\delta_\pi(4m^2_K)>180\degrees$ is obtained if  replacing 
$\alpha_1\to2.28-0.04$, i.e., moving only $\tfrac{1}{2}\,\sigma$ off the central value in the 
fits of Hyams et al.,\ref{4} Eq.~(3.2) here. Variation within errors of their  
parameters  should have taken 
DGL, ACCGL to a Type~II solution and,  therefore,
 their error for $\langle r^2_S\rangle$ should have comprised the value 
found with a Type~II solution. 
With a complete error analysis DGL, ACCGL should have got\fnote{Note
 that the converse is not true, in the sense that we do {\sl not} 
have to enlarge the errors to encompass the DGL number: 
while it is true that the {\sl parametrization} (3.1), (3.2) is compatible 
with both a solution of Type~I and one of Type~II, we have shown 
in \sect~5 that the {\sl experimental} data 
point clearly to $\delta_\pi(4m^2_K)>180\degrees$, 
hence a solution of Type~II, that 
SU(3) ch.p.t. calculations also indicate a solution of Type~II and, finally,
 in \sect~6, we have argued that only a solution of Type~II 
is compatible with the asymptotic behaviour 
indicated by perturbative QCD.} 
$\langle r^2_S\rangle=0.61_{-0.04}^{+0.21}\;{\rm fm}^2.$

For a Type~II solution, on the other hand, the value of 
 $\delta_\pi(4m^2_K)>\pi$ agrees with what experiment indicates; the phases 
 $\phi_\pi(s)$, 
$\delta^{(+)}(s)$ and $\delta_S(s)$ are continuous both in $s$
 and when the inelasticity goes to zero; and 
the phase $\delta_S(s)$ agrees well with 
what perturbative QCD suggests at large~$s$. 
We conclude that a situation of Type~II is by far the more likely to be correct, 
 thus confirming the validity of the approximations in ref.~2; 
in particular, the estimate
$$\langle r^2_S\rangle=0.75\pm0.07\,{\rm fm}^2.
\equn{(7.2)}$$

A last question is whether one can improve on the evaluation in 
ref.~2. This is very unlikely, for the 
contribution of the region $4m^2_K\leq s\leq2\,\gev$, \equn{(1.7)}.  
First of all, the incompatibility of the central values for $\eta$ 
 in analyses based on $\pi\pi\to\pi\pi$ scattering\ref{4,10,11} 
with what one finds in $\pi\pi\to\bar{K}K$ experiments,\ref{5} 
implies that the phase $\delta_\pi$ obtained from $\pi\pi\to\pi\pi$ scattering
must be biased. And, secondly, to find the eigenphases 
$\delta^{(\pm)}$ and mixing angle $\theta$ which are necessary 
to disentangle the form factors $F_{S,\pi}$, $F_{S,K}$ [cf.~Eq.~(2.6)], 
 one requires, as 
discussed in detail in ref.~7,  
experimental measurements of the {\sl three} reactions
$\pi\pi\to\pi\pi,\;\pi\pi\to\bar{K}K,\;\bar{K}K\to\bar{K}K$. 
Failing this, we are only left with approximate evaluations, 
like those in ref.~2.

\vfill\eject
\brochuresection{Acknowledgments}

\noindent
I am grateful to CICYT, Spain, and to INTAS, for financial support. 
Discussions with J.~R.~Pel\'aez have been very useful.

\booksection{References}

\item{1 }{DGL: Donoghue, J. F., Gasser, J.,  and Leutwyler, H., {\sl Nucl. Phys.}, 
{\bf B343}, 341 (1990), slightly 
improved in Colangelo,~G., Gasser,~J.,  and Leutwyler,~H., {\sl Nucl. Phys.}, 
{\bf B603}, 125 (2001).}
\item{2 }{Yndur\'ain, F. J., {\sl Phys. Lett.} {\bf B578}, 99 (2004) and
 (E), {\bf B586}, 439 (2004).}
\item{3 }{ACCGL: Ananthanarayan, B., et al., {\sl Phys. Lett.} {\bf B602}, 218 (2004).}
\item{4 }{Hyams, B., et al., {\sl Nucl. Phys.} {\bf B64}, 134 (1973).}
\item{5 }{Wetzel,~W., et al., {\sl Nucl. Phys.} {\bf B115}, 208 (1976);
Polychromatos,~V.~A., et al.,  {\sl Phys. Rev.} {\bf D19}, 1317 (1979);
Cohen, ~D. et al., {\sl Phys. Rev.} {\bf D22}, 2595 (1980); 
Etkin,~E. et al.,  {\sl Phys. Rev.} {\bf D25}, 1786 (1982).}
\item{6 }{Pel\'aez, J. R., and Yndur\'ain, F. J.,  FTUAM 04-14 (hep-ph/0411334), 
to appear in Phys.~Rev.~D.}
\item{7 }{Yndur\'ain, F. J.,
 {\sl  Nucl. Phys.}, {\bf B88}, 316 (1975).}
\item{8 }{Protopopescu, S. D., et al., {\sl Phys Rev.} {\bf D7}, 1279, (1973).}
\item{9 }{Au,~K.~L., Morgan,~D., and Pennington,~M.~R.,  {\sl Phys Rev.} {\bf D35}, 1633, (1987).}
\item{10}{Hyams, B., et al., {\sl Nucl. Phys.} {\bf B100}, 205 (1975).}
\item{11}{Grayer, G., et al.,  {\sl Nucl. Phys.}  {\bf B75}, 189, (1974); 
 Kami\'nski, R., Lesniak, L, and Rybicki, K., {\sl Z.~Phys.} {\bf C74}, 79 (1997) and 
{\sl Eur. Phys. J. direct} {\bf C4}, 4 (2002).}
\item{12}{Oller, J. A., Oset, E., and Pel\'aez,~J.~R.,   {\sl Phys. Rev.} 
{\bf D59}, 074001 (1999); G\'omez-Nicola,~A., and  and Pel\'aez,~J.~R., 
 {\sl Phys. Rev.} {\bf D65}, 054009  (2002). See also Oller, J. A. and Oset, E., {\sl Phys. Rev.} 
{\bf D60}, 074023 (1999)}
\item{13}{Brodsky, S. J., and Lepage, G. P., {\sl Phys. Rev.} {\bf D22}, 2157 (1980); 
 Brodsky,~S.~J., Frishman,~Y., Lepage, G. P., and Sachrajda, C. T., 
{\sl Phys. Lett.} {\bf 91B}, 239  (1973); and 
Efremov, A. V., and Radyushin, A. V.,  {\sl Riv. Nouvo Cimento} {\bf 3}, No. 2 (1980b),
as developed in grater detail in Yndur\'ain,~F.~J., {\sl The Theory of 
Quark and Gluon Interactions}, Springer, Berlin 1999 (\sect~5.7), which we follow here. }
\item{14}{Aubin, C., et al. MILC Collaboration (hep-lat/0407028).}
\item{15}{Yndur\'ain, F. J., FTUAM 03-14
(hep-ph/0310206, December 2004 version).}
\item{16}{Moussallam, B., {\sl Eur. Phys. J.} {\bf C14}, 111 (2000).}

\bye